\documentclass{article}

\usepackage{amssymb}
\usepackage{amsmath}
\usepackage{verbatim}

\newtheorem{theorem}{Theorem}
\newtheorem{definition}{Definition}
\newtheorem{lemma}{Lemma}

\newtheorem{Property}{Combinatorial Property}

\newcommand{\poly}{\mathrm{poly}}
\newcommand{\pr}{\mathrm{Prob}}
\newcommand{\om}{\mathrm{\Omega}}

\begin{document}

\title{Pseudo-random graphs and  bit probe schemes with one-sided  error
\thanks{Supported in part by grants
ANR EMC ANR-09-BLAN-0164-01 and NAFIT ANR-08-EMER-008-01.}}
\author{Andrei Romashchenko\\
CNRS (France)\  \&\  IITP of RAS (Russia, Moscow)}
\maketitle

\begin{abstract}
We study probabilistic bit-probe schemes for  the membership problem. Given
a set $A$ of at most $n$ elements from the universe of size $m$ we organize
such a structure that queries of type ``$x\in A$?'' can be answered very quickly.


H.~Buhrman, P.B.~Miltersen, J.~Radhakrishnan, and S.~Venkatesh proposed
a  bit-probe scheme based on expanders.  Their scheme needs space of 
$O(n\log m)$ bits. The scheme has a randomized algorithm processing
queries; it needs to read \emph{only one} randomly chosen
bit from the memory to answer a query. For every $x$ the answer is correct with 
high probability (with two-sided errors). 


In this paper we show that  for the same problem there exists a bit-probe scheme
with \emph{one-sided} error that needs space of $O(n\log^2 m+\poly(\log m))$ bits. The difference
with the model of Buhrman, Miltersen, Radhakrishnan, and Venkatesh
is that we consider a bit-probe scheme with an auxiliary word. This means that in our
scheme  the memory is split into two parts of different size: the main storage of
$O(n\log^2 m)$ bits and a short  word of $\log^{O(1)}m$  bits that is pre-computed 
once for the stored set $A$ and ``cached''. To answer a query ``$x\in A$?''
we  allow to read the whole cached word and only one bit from the main storage.
For some reasonable values of parameters (e.g., for $\poly(\log m) \ll n \ll m$) 
our space bound is better than what can be achieved by any scheme without  cached data
(the lower bound  $\om(\frac{n^2\log m}{\log n})$  was proven in \cite{bmrv}). 
%
%
We obtain a slightly weaker result (space of size $n^{1+\delta}\poly(\log m)$ bits 
and two bit  probes for every query) for a scheme that is effectively encodable.


Our construction is based on the idea of naive derandomization, 
which is of independent  interest.  First we prove that 
a random combinatorial object (a graph) has the required properties, and then
show that such a graph can be obtained as an outcome of a
pseudo-random bits generator. Thus, a suitable graph can be specified
by a short seed of a PRG, and we can put an appropriate value of the
seed  into the cache memory of the  scheme.

\end{abstract}

\section{Introduction.}

We investigate the static version of the membership problem. The aim is 
to represent a set $A\subset\{1,\ldots,m\}$ by some data structure 
so that queries ``$x\in A$?'' can be easily replied.  We are interested in
the case when the number of elements in the set $n=|A|$ is much less than size $m$ 
of the universe (e.g., $n=\mathrm{exp}\{\poly(\log \log m)\}$ or $n=m^{0.01}$).

In practice, many different data structures are used to represent
sets:  simple arrays, different variants of  height-balanced trees, 
hash tables, etc. The simplest solution is just an array of $m$ bits;
to answer a query ``$x\in A$?'' we need to read a single bit from
the memory (the $x$-th bit in the data storage is equal to $1$ if $x\in A$). 
However,  the size of this data structure is  excessive:
it requires $m$ bits of memory, while there exist only  
$\binom{m}{n}=2^{\Theta(n\log m)}$ different subsets $A$ of size
$n$ in the $m$-elements universe.

Nowadays the standard practical solution for the membership problem 
is  a  more complex data structure 
proposed by Fredman, Koml\'os, and Szemer\'edi~\cite{fks}. 
This scheme is based on perfect  hashing;
a set is represented as a table of $O(n)$ words (hash-values) of size $\log m$ bits,
and a query  ``$x\in A$?'' requires to read $O(1)$ words from the memory.
The space complexity of this construction is quite close to the trivial
lower bound  $\om(\log \binom{m}{n})$. The asymptotic of the space complexity
of this scheme was further improved 
in \cite{improved-hasing-1,improved-hasing-2,improved-hasing-3}.
Similar space complexity was achieved in \emph{dynamic} data structures,
which support fast update of the set stored in the database
(see, e.g., analysis of the cuckoo hashing scheme in \cite{cuckoo1,cuckoo2}). 
A subtle analysis of the space and bit-probe complexity for the
membership  problems was given also in \cite{pagh-stoc}
(in particular, \cite{pagh-stoc} suggested a membership scheme based on bounded 
concentrator graphs).
Note that all these schemes require to
read  from the memory $O(\log m)$ bits  to answer each query.

Another popular practical solution is Bloom's filter \cite{bloom}.
This data structure requires only $O(n)$ bits, whatever is the size 
of the universe; to answer a query we need to read $O(1)$ bits from the memory.
The drawback of this method is that we can get false answers to
some queries.  Only false positives answers are possible
(for some $x\not\in A$ Bloom's filter answers ``yes''), but false negatives are not. 
When this technique is used in practice, it is believed that  
for a ``typical'' set $A$ the fraction of false answers should be small. However,
in many applications we cannot fix \emph{a priori} any reasonable probability
distribution on the family of all sets $A$ and on the space of possible queries.

An interesting alternative approach was suggested by Harry Buhrman, Peter Bro Miltersen, 
Jaikumar Radhakrishnan, and  Venkatesh Srinivasan \cite{bmrv}. 
They introduced  randomness into the query processing algorithm. That is, 
the data structure remains static (it is deterministically defined for each set $A$), 
but when a query is processed, we a toss coins and read randomly chosen bit
from the memory. In this model,  we allow to return
a wrong answer with some small probability. Notice the sharp difference
with the Bloom's filter: Now  we must correctly reply to the query ``$x\in A$?'' 
with probability close to $1$ for  \emph{each} $x$.

Buhrman, Miltersen,  Radhakrishnan, and Venkatesh investigated
both two-sided and one-sided errors. In this paper we will concentrate
mostly on one-sided errors: if $x\in A$, then the answer must be always correct, 
and if  $x\not\in A$, then a small probability of error is allowed.

Recall that a trivial information-theoretic bound shows that the size of the structure representing
a set $A$ cannot be less that $\log \binom{m}{n}=\om(n\log m)$ bits. Surprisingly,
this bound can be achieved if we allow two-sided errors and use only \emph{single bit} probe
for each query. This  result was proven in~\cite{bmrv}. We refer to the scheme proposed their
as the   BMRV-scheme:
\begin{theorem}[two-sided BMRV-scheme, \cite{bmrv}]\label{main-th-two-side}
For any $\varepsilon>0$ there is a scheme for storing subsets $A$ of size at most
$n$ of a universe of size $m$ using $O(\frac{n}{\varepsilon^2}\log m)$
bits so that any membership query ``Is $x\in A$?''
can be answered with error probability less than $\varepsilon$ by a           
randomized algorithm which probes the memory at just one
location determined by its coin tosses and the query element $x$.
\end{theorem}
The size of the memory achieved in this theorem is only a constant factor greater
than the best possible. In fact,
the trivial lower bound $\log \binom{m}{n}$ can be improved: the less
is the probability of an error of the scheme, the more memory we need. 
\begin{theorem}[lower bound, \cite{bmrv}]\label{lower-bound} 
\textup(a\textup) For any $\varepsilon>0$ and $\frac{n}{\varepsilon}<m^{1/3}$, any 
$\varepsilon$-error randomized scheme which answers queries using one
bitprobe must use space $\om(\frac{n}{\varepsilon\log1/\varepsilon}\log m)$.

\textup(b\textup) 
Any scheme with \emph{one-sided} error $\varepsilon$ that answers queries using
at most one bitprobe must use $\om(\frac{n^2}{\varepsilon^2\log (n/\varepsilon)}\log m)$
bits of storage. 
\end{theorem}
Note that for one-sided error schemes the known lower bound is much stronger.
Part (b) of the theorem above implies that
we cannot achieve  the size of space  $O(n\log m)$
with a \emph{one probe} scheme
and one-sided error. However we can get very close if we allow 
$O(1)$ probes instead of a single probe:
\begin{theorem}[one-sided BMRV-scheme, \cite{bmrv}]\label{main-th-one-side}
Fix any  $\delta>0$. There exists a constant $t$ such that the following
holds: There is a one-sided $\frac13$-error randomized scheme that uses space
$O(n^{1+\delta}\log m)$ and answers membership queries with at most $t$ probes.
\end{theorem}

The constructions in~\cite{bmrv} is not explicit: given the list of elements $A$, 
the corresponding  scheme  is constructed  (with some brute force search) in time 
$2^{\poly(m)}$. Moreover, each membership query requires exponential 
in $m$ computations.

The crucial element of the constructions in Theorem~\ref{main-th-two-side} is an unbalanced 
expander graph.  Existence of a graph with required parameters was proven in~\cite{bmrv}
probabilistically. We know that such a graph exists and we can find it by brute force search,
 but we do not know how to construct it explicitly.
In case  we have an \emph{effective} construction of an expander with good parameters,
we will get a practical variant of the BMRV-scheme. This scheme could be also generalized
to build  more complex data structures (see \cite{pagh-one-probe} for a construction
of a \emph{dictionary} data structure based on the BMRV-scheme and some explicit expanders).

Since Bassalygo and Pinsker defined expanders~\cite{expander-def-p,expander-def}, many explicit 
(and poly-time computable) constructions of expander graphs were discovered, 
see a survey~\cite{expanders-survey}. However, most of the known constructions are based on the
spectral technique that is not suitable to get an expander of degree $d$ with an expansion parameter 
greater than $d/2$, see \cite{kah}. This is not enough for the construction used in the proof
of Theorem~\ref{main-th-two-side} in \cite{bmrv}; we need there a graph with expansion parameter 
close to $d$.

There are only very few effective constructions of  unbalanced graph with 
large expansion parameter. One of the known constructions was suggested by 
Capalbo \emph{et al} in~\cite{capalbo}; its parameters are  close to the optimal values if the size
of the right part of the graph is  \emph{constant times} less than the size of the left part
of the graph. However, in the BMRV-scheme we need a very unbalanced expander,
i.e., a graph where the right part of
the graph is much less than the left part; so, the technique from~\cite{capalbo} seems
to be not suitable here. 
Some explicit version of the BMRV-scheme was suggested in~\cite{ta-schma}  
(this construction involves  Trevisan's extractor; note that Trevisan's 
extractor is known to be a good highly unbalanced expander, \cite{ta-shma-umans-zuckerman}). 
The best known  explicit construction of 
a highly unbalanced expander graph was presented in~\cite{parvaresh-vardy}. 
It is based on the Parvaresh--Vardy code with an efficient list decoding. 
Thanks to the special structure if this expander, it enjoys some nice  property of
effective decoding. Using this technique, the following variant of Theorem~\ref{main-th-two-side}
can be proven:

\begin{theorem}[\cite{parvaresh-vardy}]\label{main-th-parvaresh-vardy}
For any $\delta>0$ there exists a scheme for storing subset $A$ of size at most
$n$ of a universe of size $m$ using $n^{1+\delta}\cdot \poly(\log m)$
bits so that any membership query 
can be answered with error probability less than $\varepsilon$ by a           
randomized algorithm which probes the memory at  one
location determined by its coin tosses and the query element $x$.

Given the list of elements $A$, the corresponding storing scheme can be constructed
in time $\poly(\log m, n)$. When the storing scheme is constructed, a query for an 
element $x$ can be calculated in time $\poly(\log m)$.
\end{theorem}

In Theorems~\ref{main-th-two-side},~\ref{main-th-one-side},~\ref{main-th-parvaresh-vardy},
a set $A$ is encoded into a bit string, and when we want to know if $x\in A$, we just
read from this string one  
randomly chosen bit (or $O(1)$ bits in Theorem~\ref{main-th-one-side}). 
The obtained information
is enough to decide whether $x$ is an element of the set. Let us notice that  in all
these computations we implicitly use  more information than just a single bit extracted 
from the memory. To make a query to the
scheme and to process the retrieved bit, we need to know the parameters of the
scheme: the size $n$ of the set $A$, the size $m$ of the universe, 
and the allowed error probability $\varepsilon$. This auxiliary
information is very short (it takes only $\log (m/\varepsilon)$ bits), and it does not depend
on the stored set $A$. We assume that this information
is somehow hardwired into the bitprobe scheme (we shall  say that this information
is \emph{cached} in advance by the algorithms that processes  queries).

\medskip

In this paper we consider a more liberal model, where some small  
information \emph{cached} by the scheme
can depend not only on $n, m$, and $\varepsilon$, but also on the set $A$. 
Technically,  the data stored in our scheme 
consists of two parts of different size: a small cached string $C$ of length
$\poly(\log m)$, and a long bit string $B$ of length $n\cdot \poly\log(m)$. 
Both these strings are prepared for a given set $A$ of  $n$
elements (in the universe of size $m$). When we need to answer a query ``$x\in A$?'',
we use $C$ to compute probabilistically a position in $B$ and read there one
bit. This is enough to answer whether $x$ is an element of $A$, with a small one-sided
error:

\begin{theorem}\label{main-th-1}
Fix any constant $\varepsilon>0$. There exists 
a one-sided $\varepsilon$-error randomized scheme that includes
a string $B$ of length $O(n\log^2 m)$ and
an auxiliary word $C$ of length $\poly(\log m)$.
We can answer membership queries ``$x\in A$?'' with one bit probe to $B$.
For $x\in A$ the answer is always correct; for each $x\not\in A$ probability of error is
less than $\varepsilon$.

The position of the bit probed in $A$ is computed from $x$ and the auxiliary word $C$
in time $\poly(\log m)$.
\end{theorem}
\textbf{Remark~1:} Schemes with `cached' auxiliary information that depends on $A$ 
(not only on its size $n=|A|$ and the size $m$ of the universe) make sense only 
if the cached information is
very small. Indeed, if the size of the cached data is about $\log\binom{m}{n}$ bits, 
then we can put there 
the list of all elements of $A$, so the problem becomes trivial. 
Since in our construction we need 
cached information of size $\poly(\log m)$ bits, the result is interesting when
$\poly(\log m)\ll n\ll m$, e.g., for $n=\mathrm{exp}\{\poly(\log\log n)\}$.
Note that by Theorem~\ref{lower-bound} the space size $O(n\log^2 m)$ with one-sided
error cannot be achieved by  any schemes without  cached auxiliary information 
that depends on $A$.

\medskip
\noindent
\textbf{Remark~2:} The model of data structures with cached memory looks
useful for practical applications. Indeed,  most computer systems
contain some hierarchy of memory levels: CPU registers
and several levels of processor caches, then  random access memory,
flash memory, magnetic disks, remote network-accessible drives,
etc. Each next level of memory is cheaper but slower. 
So, it is interesting to investigate the tradeoff between expensive and fast
local memory and cheap and slow external memory. 
There is a rich  literature on algorithms with \emph{external memory}, see,
e.g, surveys \cite{vitter,vitter2}.
Thus, the idea of splitting the data structure into `cached' and `remote'
parts is very natural and quite common in computer science. 
However, tradeoff between local and external memory is typically
studied for \emph{dynamic} data structures.
The same time, it is not obvious that fast cache memory of negligible size 
can help to process queries in a \emph{static} data structure. Since
a small cache `knows' virtually nothing about most objects in the database,
at first sight it seems to be useless. However, Theorem~\ref{main-th-1} 
shows that even a very small  cache can be surprisingly efficient.

\medskip
\noindent
\textbf{Remark~3:}  In the proof of Theorem~\ref{main-th-1}
we derandomize a probabilistic proof of existence of some kind of expander graphs.
In many papers derandomization of probabilistic arguments
involves  highly sophisticated \emph{ad-hoc} techniques. 
But we do derandomization in rather naive and straightforward way: 
take a value of a suitable
pseudo-random bits generator and check that with high probability
(i.e., for most values of the seed) a pseudo-random objects enjoys the required property. 
In fact, we observe
that several types of generators fit our construction. Since the required property of
a graph can be tested in $\mathrm{AC}^0$, we can use the classic
Nisan--Wigderson generator or 
(thanks to the recent result of Braverman~\cite{braverman})
any polylog-independent function.  
Also the required property of a pseudo-random graph can be tested 
by a machine with logarithmic space. Hence, we can use Nisan's generator \cite{nisan}.
Our idea of employing pseudo-random structures is quite similar to the construction
of pseudo-random hash-functions in~\cite{cuckoo}.
We stress that  we do not need any unproven assumptions to construct all these generators.
 

\medskip

In Theorem~\ref{main-th-1} we construct  a scheme such that decoding is effective:
when the scheme is prepared, we can answer queries ``$x\in A$?'' in time polynomial
in $\log m$. However the encoding (preparing the database and the auxiliary word
for a given set $A$)
runs in expected time $\poly(m)$. We assume that $n\ll m$, and the time polynomial in
$m$ seems to be too long. It is natural
to require that encoding of the scheme runs in time $\poly(n,\log m)$ (i.e., polynomial
in the size of the encoded set $A$ and the size of an index of each element in the universe). 
The next theorem claims
that the encoding time can be reduced if we slightly increase the  space of the scheme:

\begin{theorem}\label{main-th-2}
The scheme from theorem~\ref{main-th-1} can be made effectively encodable
in the following sense. Fix any constants $\varepsilon,\delta>0$. There exists 
randomized scheme that includes a bit string $B$ of length $n^{1+\delta}\poly(\log m)$ 
and an auxiliary word $C$ of length $\poly(\log m)$.
We can answer membership queries ``$x\in A$?'' with two bits probe to $B$.
For $x\in A$ the answer is always correct; for $x\not\in A$ probability of error is
less than $\varepsilon$.

The position of the bit probed in $A$ is computed by $x$ and the auxiliary word $C$
in time $\poly\log(m)$. Given $A$, the entire scheme \textup(the string $B$ and 
the word $C$\textup) can be computed probabilistically in average time
$\poly(n,\log m)$.
\end{theorem}

The rest of the paper is organized as follows. In Section~2 we remind the main ideas
in the BMRV-scheme. We prove Theorem~\ref{main-th-1} in Section~3, and 
Theorem~\ref{main-th-2} in Section~4. In Section~\ref{section-lower-bound}
we show that the proof of Theorem~\ref{lower-bound}~(a)
\emph{mutatis mutandis} can be applied to our model with small cached memory.
In Conclusion we discuss some open questions.

\section{How BMRV-scheme works.}\label{bmrv-section}
Let us explain the main ideas of the proof of Theorem~\ref{main-th-two-side}  in \cite{bmrv}.
The construction is based on highly unbalanced \emph{expanders}.

\begin{definition}
A bipartite graph $\mathcal{G}=(L,R,E)$ \textup(with left part $L$, right part $R$ and set of edges $E$\textup)
is called $(m,s,d, k,\delta)$-expander  if
$L$ consists of $m$ vertices,  $R$ consists of $s$
vertices, degree of each vertex in $L$ is equal to $d$, and for each subset of
vertices $A\subset L$ of size at most $k$  the number of neighbors is 
at least $(1-\delta)d|A|$. 
\end{definition}
We use a standard notation: for a vertex $v$ we denote by $\Gamma(v)$  the set of 
its neighbors; for a set of vertices $A$ we denote by $\Gamma(A)$
the set of neighbors of $A$, i.e., $\Gamma(A) = \cup_{v\in A} \Gamma(v)$.
So, the definition of expanders claims that for all small enough sets $A$
of vertices in the left part of the graph, $|\Gamma(A)|\ge (1-\delta)d|A|$
(the maximal size of $|\Gamma(A)|$ is obviously $d|A|$, since degrees of all
vertices on the left are equal to $d$).
The argument below is based on the following combinatorial 
property of  an expander:
\begin{lemma}[see \cite{capalbo}]\label{expander-lemma}
Let $\varepsilon$ be a positive number, and 
 $\mathcal{G}$ be an $(m,s,d,k,\delta)$-expander with $\delta\le \varepsilon/4$.
Then for every subset $A\subset L$ such that $|A|\le k/2$, 
the number of vertices $x\in L\setminus A$
such that  
 $$|\Gamma(x)\cap \Gamma(A)|\ge \varepsilon  d$$ 
 is not greater than $|A|/2$.
\end{lemma}
Let $\mathcal{G}$ be a $(m,s,d,k,\delta)$-expander with $\delta < \varepsilon/4$. 
The storage scheme is defined as follows. We identify  a
set $A\subset\{1,\ldots,m\}$ of size $n$  ($n\le k/2$)
with a subset of vertices
in the left part of the graph. We will represent it by some 
labeling (by ones and zeros) on the vertices of the right part of the graph. 
We do it  in such a way that the vast majority (at least $(1-\varepsilon)d$)
of neighbors  of each vertex $v$ from the left part of the graph correctly 
indicate whether $v\in A$. More precisely, if $v\in A$ then at least $(1-\varepsilon)d$
of its neighbors in $R$ are labeled by $1$; if  $v\in L\setminus A$ then at least 
$(1-\varepsilon)d$ of its neighbors in $R$ are labeled by $0$.
Thus, querying a random neighbor of $v$  will 
return the right answer  with probability $>1-\varepsilon$.

It remains to explain why such a labeling exists. In fact, it can
be constructed by a simple greedy algorithm. 
First, we label all neighbors of $A$  by $1$, and the other vertices  on
the left by $0$. This labeling classifies correctly all vertices in $A$.
But it can misclassify some vertices outside $A$: 
some  vertices in $L\setminus A$ can have too many
(more than $\varepsilon d$) neighbors labeled by $1$. Denote by $B$ the set
of  all these ``erroneous'' vertices. We relabel all their neighbors,
i.e., all vertices in $\Gamma(B)$ to $0$. This
fixes the problem with vertices outside $A$, but it can 
create problems with some vertices in $A$.  We take the set of all 
vertices in $A$ that became erroneous (i.e., vertices in $A$
that have at least $\varepsilon d$
neighbors in $\Gamma(B)$), and denote this set of vertices by $A'$.
Then, we relabel all $\Gamma(A')$ to $1$. This operation create
new problems in some set of vertices $B'\subset B$, we relabel $\Gamma(B')$ to $0$,
etc. In this iterative procedure we get a sequence of sets 
 $$
 A \supset A' \supset A'' \supset\ldots
 $$
 whose neighbors are relabeled to $1$ on steps $1,3,5,\ldots$ of the algorithm, and
 $$
 B \supset B' \supset B'' \supset\ldots
 $$ 
 whose neighbors are relabeled to $0$ on iterations $2,4,6,\ldots$ respectively.
Lemma~\ref{expander-lemma} guarantees that 
the number of the erroneous vertices on each iteration reduces by 
a factor of $2$ ($|B| \le |A|/2$, $|A'|\le |B|/2$, etc.). 
Hence, in $\log m$ steps the procedure terminates.

To organize a storing scheme (and to estimate its size) for a set $A$ of size $n$
in the universe  of size $m$, we should construct an 
$(m,s,d, k=2n,\delta=\varepsilon/4)$-expander.
Parameters $m,k,\delta$ of the graph are determined directly by the parameters 
of the desired scheme
(by the size of $A$ and the universe and the allowed error probability $\varepsilon$).
 We want to minimize the size of the left part of the graph $s$, which is the size of
 the stored data. Existence of expanders with good parameters can be proven
 by probabilistic arguments:
  \begin{lemma}[\cite{bmrv}]\label{expander-existence-lemma}
For all integers $m,n$ and real $\varepsilon>0$
there exists an $(m, s= O(\frac{n\log m}{\varepsilon^2}), d=\frac{\log m}{\varepsilon}, n, \varepsilon)$-expander.
Moreover, the vast majority of bipartite graphs with $n$ vertices on the left, $s=\frac{100n\log m}{\varepsilon^2}$
vertices on the right, and degree $d= \frac{\log m}{\varepsilon}$ at all vertices on the left are such expanders.
   \end{lemma}
 Given the parameters $m, n, \varepsilon$, we can find an 
 $(m, O(\frac{n\log m}{\varepsilon^2}), \frac{\log m}{\varepsilon}, n, \varepsilon)$-expander
 by brute force search. This can be done by a deterministic algorithm in time $2^{\poly(m/\varepsilon)}$.
 Hence, we can construct the bit-probe structure defined above in exponential time. Moreover,
 when the structure is constructed and we want to answer a query ``$x\in A$?'', we need to read only
 one bit from the stored bit string. But to select the position of  this bit we need again to reconstruct
 the expander graph, which requires exponential computations. We could keep the structure
 of the computed graph in ``cache'' (compute the graph once, and then re-use it every time a
 new query should be answered). However, in this case 
 the size of the ``cached data'' (the size of the graph) becomes 
 much greater than $m$, which makes the bit-probe scheme meaningless (it is cheaper
 to store $A$ as a trivial $m$-bits array).
 
 In~\cite{parvaresh-vardy} a nice and very powerful explicit construction of expanders was suggested:
  \begin{theorem}[\cite{parvaresh-vardy}]\label{th-parvaresh-vardy-expander}
Fix an $\varepsilon>0$ and $\delta>0$.
For all integers $m,n$ 
there exists an explicit
$(m, s= n^{1+\delta}\cdot \poly(\log m), d=\poly(\log m), n, \varepsilon)$-expander
such that for an index of a vertex $v$ from the left part
\textup(a binary representation of an integer between $1$ and $m$\textup)  and an index of 
an outgoing edge \textup(a binary representation of an integer between $1$ and $d$\textup), 
the corresponding
neighbor on the right part of the graph 
\textup(an integer between $1$ and $s$\textup) can be computed 
in time polynomial in $\log m$.

Also, the following effective decoding algorithm exists. Given a set of vertices $T$ from the right part
of the graph, we can compute the list of vertices in the left part of the graph that have at least
$(4\varepsilon d)$ neighbors in $T$, i.e., 
 $$
 S = \{ v\ :\ |\Gamma(v)\cap T|\ge 4\varepsilon d \},
 $$
in time $\poly(|S|, n, \log m)$. 
   \end{theorem}
Theorem~\ref{main-th-parvaresh-vardy} is proven by plugging the expander from 
Theorem~\ref{th-parvaresh-vardy-expander} in the general scheme explained above,
see details in \cite{parvaresh-vardy}.

\section{Proof of Theorem~\ref{main-th-1}.}

\subsection{Refinement of the property of $\varepsilon$-reduction.}

The construction of a bit-probe scheme for a set $A$ of size $n$ in the $m$-elements universe
(with probability of an error bounded by some $\varepsilon$) explained in the previous section 
involves an $(m,s,d,k,\delta)$-expander with $s=O(\frac{n}{\varepsilon^2}\log m)$ and 
$d=O(\frac{1}{\varepsilon}\log m)$. Such a graph contains $dm$ edges (degree of each vertex on
the left is $d$).  The list of all its edges can be specified by a string of $dm\log s$ bits: 
we sort all edges by their left ends, and  specify for each edge its right end. Denote the size
of the description of this graph by  $N=dm\log s$. 

In what follows we will assume that  number $s$ is a power of $2$ (this will increase the  parameters of 
the graph only by a factor at most $2$). So, we may assume that every string of $N(m,s,d)=dm\log s$
bits specifies a bipartite graph  with $m$ vertices on the left, $s$ vertices on the right and degree $d$
on the left.
Lemma~\ref{expander-existence-lemma} claims that most  of these bits string of length $N$ describe an
$(m,s,d,k,\delta)$-expander.   
By  Lemma~\ref{expander-lemma}, if a graph is an expander with these parameters, 
then for $\varepsilon=4\delta$ and 
for every set $A\subset L$ of size less than $k/2$
 the following \emph{reduction property} holds:
\begin{Property}[$\varepsilon$-reduction property]\label{reduction-property}
For every subset $A\subset L$ such that $|A|\le k/2$,  
the number of vertices $x\in L\setminus A$ such that  
$$|\Gamma(x)\cap \Gamma(A)|\ge \varepsilon d$$ 
is not greater than $|A|/2$.
\end{Property}
This property was the main ingredient of the BMRV-scheme. In our bit-probe scheme we will need 
another  variant of Property~\ref{reduction-property}:
\begin{Property}[strong $\varepsilon$-reduction]\label{strong-reduction-property}
Let $\mathcal{G}=(L,R,E)$ be a bipartite  graph, and $A\subset L$ be a subset of vertices
from the left part.
We say that the \emph{strong $\varepsilon$-reduction property} holds for $A$ in this graph if
for all $x\in L\setminus A$
 $$|\Gamma(x)\cap \Gamma(A)|\le \varepsilon d.$$ 
\end{Property}
\begin{lemma}\label{strong-reduction}
Fix an $\varepsilon>0$. For all integers $m,n$, for every $A\subset\{1,\ldots,m\}$ of size $n$
there exists a bi-partite graph $\mathcal{G}=(L,R,E)$ such that 
 \begin{itemize}
\item $|L|=m$ \textup(the size of the left part\textup);
\item $|R|=2 d^2n = O(n\log^2m)$ \textup(the size of the right part\textup);
\item degree of each vertex in the left part is $d=\frac{2\log m}{\varepsilon} = O(\log m)$;
\item the property of strong $\varepsilon$-reduction holds for the set $A$ 
\textup(we identify it with a subset of vertices in left part of the graph\textup).
 \end{itemize}
Moreover,  the property of strong $\varepsilon$-reduction for $A$ holds for the majority
of graphs with the parameters specified above.
\end{lemma}
The order of quantifiers is important here: we do not claim that in a random graph
the strong $\varepsilon$-reduction property holds for all $A$; we say only that for
every $A$ the strong $\varepsilon$-reduction is true in a random graph.

\medskip

\noindent
\textbf{Proof of lemma:}
Let $v$ be any vertex in $L\setminus A$. We estimate probability that at least
$\varepsilon d$ neighbors of $x$ are at the same time neighbors of $A$
(assuming that all edges are chosen at random independently). 
There are $\binom{d}{\varepsilon d}$ choices of $\varepsilon d$ vertices among
all neighbors of $v$. Hence, 
 $$
 \pr[|\Gamma(v)\cap \Gamma(A) |\ge \varepsilon d] \le
 \binom{d}{\varepsilon d}  \cdot \left(\frac{|\Gamma(A)|}{|R|}\right)^{\varepsilon d} \le
  d^{\varepsilon d} \cdot \left(\frac{ dn}{2 d^2 n}\right)^{\varepsilon d} 
  = \left(\frac{ 1}{2 }\right)^{2 \log m} 
 $$
This probability is less than $1/m^2$ (for each vertex $v$). So, the expected number 
of vertices $v\in L$ such that  $|\Gamma(v)\cap \Gamma(A) |\ge \varepsilon d$, is
less than $1/m<1/2$. Hence, the strong $\varepsilon$-reduction property holds for  $A$
for more than $50\%$ of graphs.

\subsection{Testing the property of strong $\varepsilon$-reduction.}

Lemma~\ref{strong-reduction} implies that a graph with the strong reduction property
for $A$ exists. Given $A$, we can find such a graph by brute force search.
But we cannot use such a graph in our bit-probe scheme even if we
do not care about computation complexity: the choice of the graph depends on $A$,
and the size of the graph is too large to embed it  into the scheme explicitly. We need to find
a suitable graph with a short description. We will do it using pseudo-random bits generators
(`pseudo-random' graphs will be parameterized by the seed of a generator).

Property~\ref{strong-reduction-property} is a property of a graph and of a set of vertices $A$ in this 
graph. We
can interpreted it as a property of an $N$-bits string (that determines a graph)
and some $A\subset\{1,\ldots,m\}$. 
Lemma~\ref{strong-reduction} claims that  for every $A$, 
for a randomly chosen graph (a randomly chosen $N$-bits string) with high probability 
the strong reduction property is true. We want to  show that the same is true for a 
\emph{pseudo-random} graph. 
At first, we observe that the strong reduction property can be tested 
by an $\mathrm{AC}^0$ circuit (a Boolean circuit  of bounded depth, with polynomial 
number of gates \emph{and}, \emph{or} with unbounded fan-in, and \emph{negations}).

Indeed, we need to check for each vertex $v\in L\setminus A$ 
that the number of vertices in $\Gamma(v)\cap \Gamma(A)$
is not large.  For each vertex $w$ in the right part of the graph 
we can compute by an $\mathrm{AC}^0$-circuit  
whether  $w\in \Gamma(A)$:
$$
\bigvee\limits_{u\in A} \ \bigvee\limits_{i\le d} \ [\mbox{ the $i$-th neighbor of $u$ is $w$ }]
$$
(the condition in the square brackets is a statement about one edge in a graph, i.e., 
it is a conjunction of $O(\log N)$ bits and negations of bits 
from the representation of this graph).
So, for each $v\in L\setminus A$  and for $i=1,\ldots,d$
we can compute whether the $i$-th neighbor of $v$ belongs to $\Gamma(A)$:
$$
b_{v,i} = \bigvee\limits_{w\in R} \left( [\mbox{ the $i$-th neighbor of $v$ is $w$ }] \ \&\  [\ w\in \Gamma(A)\ ] \right)
$$
It remains to `count'
for each $v\in L\setminus A$ the number of neighbors in
$\Gamma(A)$ and compute the thresholds
$$
\mathrm{Th}(b_{v,1},\ldots, b_{v,d})=\left\{
\begin{array}{rl}
1,&\mbox{if }  b_{v,1}+\cdots+ b_{v,d}\ge \varepsilon d,\\
0,&\mbox{otherwise}.
\end{array}
\right.
$$ 
In $\mathrm{AC}^0$ we cannot compute thresholds with linear number of inputs 
(e.g., the majority  function is not in $\mathrm{AC}^0$, see \cite{hastad}). 
However,  we need threshold functions with only logarithmic  number of inputs. 
Such a function can be represented  by a CNF of size 
$2^{O(d)} = \poly(N)$.

Then, we combine together  these thresholds for all $v\in L\setminus A$,
and get an  $\mathrm{AC}^0$-circuit that tests the property of strong
$\varepsilon$-reduction.

\subsection{Pseudo-random graphs.}

We need to generate a pseudo-random string of $N$ bits that satisfies 
the strong $\varepsilon$-reduction property (for some fixed set $A$). We
know that (i) by Lemma~\ref{strong-reduction}, for a uniformly distributed random  
string this property is true with high probability, and 
(ii) this property can be checked in $\mathrm{AC}^0$. It remains to choose 
a pseudo-random bits generator that fools this particular $\mathrm{AC}^0$-circuit.
There exist several generators that fools such distinguishers.
Below we mention three different solutions.

\medskip

\noindent
\textbf{Remark:} We can test by an $\mathrm{AC}^0$-circuit  Property~\ref{strong-reduction-property} 
 for every fixed set $A$ but \emph{not} for all sets $A$ together.

\medskip

\noindent
\emph{The first solution: the generator of Nisan and Wigderson.}
The classic way to fool an $\mathrm{AC}^0$ circuit is the Nisan--Wigderson generator:
\begin{theorem}[Nisan--Wigderson generator, \cite{nw}]
For every constant $c$ there exists an explicit family of functions
 $$G_m:\{0, 1\}^{\poly(\log N)}\to \{0, 1\}^{N}$$
such that for  
for any family of circuits $C_N$ \textup(with $N$ inputs\textup) of polynomial in $N$ size and depth $c$, 
the difference  
$$
 \left| \pr_y[C_N(y)=1] - \pr_z[C_N(G_m(z))=1] \right|
$$ 
tends to zero \textup(faster than $1/\poly(N)$\textup).

The generator is effective: generator's value $G_m(x)$ can be computed from a given $x$
in time $\poly(\log N)$.
\end{theorem}
From this theorem and  Lemma~\ref{strong-reduction} 
it follows that for each $A\subset\{1,\ldots,m\}$ of size at most $n$, for most values of the seed 
of the Nisan--Wigderson generator $G_m$, a pseudo-random graph $G_m(x)$
satisfies  the strong $\varepsilon$-expansion property for $A$.

\medskip

\noindent
\emph{The second solution: polylog-independent strings.} M.~Braverman proves that all
polylog-independent functions fool $\mathrm{AC}^0$-circuits:
\begin{theorem}[\cite{braverman}]
Let $\mathcal{C}$ be a Boolean circuit of depth $r$ and size $M$,
$\varepsilon$ be a positive number, and
$$
 D= \left(\log \frac{M}{\varepsilon}\right)^{\kappa r^2}
$$ 
\textup(for some absolute constant $\kappa$\textup).
Then $\mathcal{C}$ cannot distinguish between the uniform distribution $U$
and any $D$-independent distribution $\mu$ on its inputs:
$$
  \left|  \pr_\mu[\mathcal{C}(x)\mbox{ accepts a $\mu$-random }x] -  
    \pr_{U}[\mathcal{C}(x)\mbox{ accepts a $U$-random }x] \right| < \varepsilon.
$$

\end{theorem}
It follows that instead of the Nisan--Wigderson generator we can take any $(\log^c n)$-independent
function (for large enough constant $c$). The standard way to generate $r$-independent bits 
(in our case we need $r=\log^c n$) is a 
polynomial of degree $r$  over a finite field of size about $N$. Seeds of this `pseudo-random bits generator'
are  coefficients of a  polynomial. Also, other (more computationally effective) constructions
of polylog-independent functions can be used. E.g.,  the construction
from \cite{siegel-89,siegel-95}  provides a family of $(\log^c n)$-independent
functions with very fast evaluation algorithm, and each function is specified by 
$\poly(\log n)$ bits (so, the size of the \emph{seed} is again poly-logarithmic).

\medskip

\noindent
\emph{The third solution: the generator of Nisan.}  
The property of strong $\varepsilon$-reduction can be tested by a Turing machine
with logarithmic working space. Technically we need a machine with 
\begin{itemize}
\item \emph{advice tape:} read-only, two-way tape, where the list of elements of $A$ is written;
\item \emph{input tape:}  read-only tape with random (or pseudo-random) bits,
with logarithmic number of passes (the machine is allowed to pass along
the input on this tape  only $O(\log N)$ times);
\item \emph{index tape:} read-only, two-way tape with logarithmic additional information;
\item \emph{work tape} that is two-way and read-write; the zone of the working tape
is restricted to $O(\log N)$.
\end{itemize}
We interpret the content of the input tape as a list of edges of a random (or pseudo-random)
graph $\mathcal{G}=(L,R)$.
The content of the index tape is understood as an index of a vertex $v\in L\setminus A$.
The machine reads the bits from the `input tape' (understood as a list of edges of a 
random graph) and checks that the vast majority of neighbors of $v$ 
does not belong to the set of
neighbors of $A$.  The machine needs to read the input tape $2d=O(\log N)$ times
(where $d$ is degree of $v$): on the first pass we find the index of the first neighbor
of $v$; on the second pass we check whether this neighbor of $v$
is incident to any vertex of $A$; then we find the second neighbor of $v$,
check whether it is is incident to any vertex of $A$, etc. The machine \emph{accepts}
the input if $|\Gamma(v)\cap\Gamma(A)|<\varepsilon d$.

We can use Nisan's  generator \cite{nisan} to fool this machine. Indeed, this checking 
procedure  fits the  general framework of \cite{sivakumar}, 
where Nisan’s generator was used to derandomize several combinatorial constructions. 
The only nonconventional feature in our argument is that the input tape is not \emph{read-once}:
we allow to read the tape with random bits  logarithmic number of 
times\footnote{The same argument can be presented in a more standard  framework,
with a read-once \emph{input tape} and an \emph{index tape} of poly-logarithmic size. However,
we believe that the argument becomes more  intuitive when we 
allow many passes on the \emph{input tape}.}.
But we  can apply  Nisan's technique for a machine that reads random bits several times.
David, Papakonstantinou, and Sidiropoulos  observed (see \cite{david}) 
that a log-space machine with  logarithmic (and even poly-logarithmic) number of passes on
the input tape is fooled by Nisan's generator  with a seed of size
$\poly(\log N)$.

\medskip

Now we are ready to prove Theorem~\ref{main-th-1}. We fix an $\varepsilon>0$ and
a set $A\subset\{1,\ldots,m\}$ of size m. Let $G_m$ be one of the
pseudo-random bits generators discussed above. For all these generators,
for most values of the seed $z$ the values $G_m(z)$ encodes a graph
such that the strong $\varepsilon$-reduction property holds for $A$. Let us
fix one of such seeds. We label by $1$ all vertices in $\Gamma(A)$ and by
$0$ all other vertices in $R$ in the graph encoded by the string $G_m(z)$.

The seed value $z$ makes the ``auxiliary word'' $C$, and the specified above
labeling of the right part of the graph makes the bit string $B$. To answer
a query ``$x\in A$?'' we take a random neighbor of $x$ in the graph and check its
label. If the label is $1$, we answer ``$x\in A$''; otherwise  ``$x\not\in A$''.

If $x\in A$, then there are no errors, since all neighbors of $A$ are labeled by $1$.
If $x\not\in A$, then probability of an error is bounded by $\varepsilon$ because
of the strong $\varepsilon$-reduction property. We can answer a query in
time $\poly(\log m)$ since the generators under consideration are 
effectively computable.

\subsection{Complexity of encoding}

The disadvantage of this construction is non-effective encoding procedure.
We know that for most seeds $z$ the corresponding graph $G_m(z)$ enjoys
the strong $\varepsilon$-reduction property. However, we need the brute force
search over all vertices $v\in L\setminus A$
(polynomial in $m$ but not in $\log m$) to check this property for any
particular seed. Thus, we have a probabilistic encoding procedure that runs in 
\emph{expected} time $\poly(m)$: we choose random seeds until
we find one suitable for the given $A$. 

In the next section we explain how to make the encoding procedure more effective 
(in expected time $\poly(n,\log m)$)
for the following price: we will need a slightly greater size of the data storage, and 
we will take $2$ bit probes instead of one at each query.

\section{Proof of theorem~\ref{main-th-2}: effective encoding.}

To obtain a scheme with effective encoding and decoding we combine two constructions:
the explicit expander from~\cite{parvaresh-vardy} and a pseudo-random graph from the 
previous section.

We fix an $n$-element set $A$ in the universe $\{1,\ldots,m\}$. Now we construct 
two bipartite graphs $\mathcal{G}_1$ and $\mathcal{G}_2$ 
that share the same left part $L=\{1,\ldots,m\}$.
The first graph is the explicit 
$(m, s= n^{1+\delta}\cdot \poly(\log m), d=\poly(\log m), n, \varepsilon)$-expander
$\mathcal{G}_1=(L,R_1,E_1)$
from~\cite{parvaresh-vardy} with an effective decoding algorithm. We do the first two steps
from the encoding procedure of the BMRV-scheme explained in Section~\ref{bmrv-section}.
At first we label all vertices in $\Gamma(A)$ by $1$ and other vertices by $0$. Denote the 
corresponding labeling  (which is a $n^{1+\delta}\cdot \poly(\log m)$-bits string)
by $B_1$. Then, we find
the list of vertices outside $A$ that have too many $1$-labeled neighbors:
 $$
  W := \{ v\in L\setminus A \ :\ |\Gamma(x)\cap \Gamma(A)|\ge \varepsilon d \}.
  $$
We do not re-label  neighbors of $W$, but we will use this set later
(to find $W$ effectively,   we need the property of effective decoding of the graph).

 Let $v\in L$ be a vertex in the left part of the graph.  There are three  different cases:
 \begin{itemize}
 \item if $v$ belongs to $A$ then all neighbors of $v$ are labeled by $1$;
 \item if $v$ does not belong to $A\cup W$, then  a random neighbor of $x$ with probability $>(1-\varepsilon)d$
is  labeled   by $0$;
\item  if $v$  belongs  to $W$, we cannot say anything certain about labels of its neighbors.
 \end{itemize}
Thus, if we take a random neighbor of $v$ and see label $0$ in $B_1$, then
we can say that this point does not belong to $A$. If we see label $1$, then a more detailed
investigation is needed. This investigation  will involve the second part of the scheme, 
which we define below.

 Now our  goal   is to distinguish between $A$ and $W$. To this end,  we take 
 a pseudo-random graph $\mathcal{G}_2=(L,R_2,E_2)$ specified by a value
 of a pseudo-random bits generator $G_m(z)$ (any of the generators discussed in the previous section
 is suitable). 
 We need a \emph{restricted
on $W$  version of the strong $\varepsilon$-reduction property}:
 
 \medskip
 \emph{
 For every $v\in W$, at most $\varepsilon d$ vertices in $\Gamma(v)$
 belong to $\Gamma(A)$.
  }
 \medskip
 
 Set $W$ is of size at most $|A|/2$ (Lemma~\ref{expander-lemma}), and it can be
 effectively computed from $A$ (effective decoding property of the graph $\mathcal{G}_1$). Hence,
 for a given $z$ we can check the property above in time $\poly(n,\log m)$. We know
 that for the majority of seeds $z$, the graph $G_m(z)$ satisfies the strong
 $\varepsilon$-reduction property, i.e., all vertices outside $A$ have at most 
 $\varepsilon d$ neighbors in $\Gamma(A)$. Though we cannot effectively
 check this general property (we cannot check it effectively for \emph{all} vertices in the universe), 
 we are able to check its restricted version (i.e., only for vertices in $W$).

Thus, in average time $\poly(n,\log m)$ we can probabilistically find some seed $z$ such 
that the restricted (on $W$) version of the strong $\varepsilon$-reduction property is true.
In the corresponding graph $\mathcal{G}_2$ we label by $1$ all vertices in $\Gamma(A)$, and by $0$
all vertices of the right part of the graph outside $\Gamma(A)$. We denote this 
labeling  (a $O(m\log^2 n)$-bits string) by $B_2$ and take it as the second part of the data storage.
The corresponding seed value $z$ is taken as  `cached' memory.

The decoding procedure works as follows. Given $x\in\{1,\ldots,m\}$, we take its random neighbor
in both constructed graphs and look at their labels (bits from  $B_1$ and $B_2$ respectively).
\begin{itemize}
\item if the first label is $0$, we say that $x\not\in A$;
\item if the first label is $1$ and the second bit is $0$ then we say that $x\not\in A$.
\item if both labels are is $1$ then we say that $x\in A$.
\end{itemize}
If $x\not\in A\cup W$, then by the definition of $W$ we know that the procedure above with probability
$>(1-\varepsilon)$  returns the correct answer. If $x\in A$, then by construction, both labels are
equal to $1$, and the procedure returns the correct answer with probability $1$. 
If $x\in W$, then we have no guarantee about labels in $B_1$; but from the restricted strong 
reducibility property it follows that with probability $>(1-\varepsilon)$ the second label is $0$. Thus,
we have one-sided error probability bounded by $\varepsilon$.

\section{A lower bound for schemes with cached memory.}\label{section-lower-bound}

In~\cite{bmrv} the lower bound $\om(n^2\log m)$ was proven for one-probe schemes with 
one-sided errors. This result cannot be applied  to schemes with small ``cached'' memory 
dependent on $A$. In fact, our scheme of size $\om(n\log m)$ with a cache of size $\poly(\log m)$ 
bits (from Theorem~\ref{main-th-1}) is below this bound for $n\gg \log^{O(1)} m$. 

On the other hand, the proof of the lower bound   $\om(\frac{n}{\varepsilon\log(1/\varepsilon)}\log m)$
(theorem~2 in~\cite{bmrv}) with minimal changes works for  schemes with cached data
if the size of the pre-computed and cached information is much less than $n\log m$:

\begin{theorem}
We consider randomized schemes 
that store sets of $n$ elements from a universe of size $m$, with two-parts
memory \textup(the cached memory of size $\poly(\log m)$ and the main storage\textup).

For all constant $\varepsilon<1$, if $\poly(\log m)\ll n \ll \sqrt[3]{m}$, 
then any such scheme with error probability $\varepsilon$ 
\textup(possibly with two-sided errors\textup) 
that answers queries using cached memory of size $\poly\log (m)$
and one bit probe to the main storage,
must use space $\om(\frac{n}{\varepsilon\log(1/\varepsilon)}\log m)$.
\end{theorem}

\textbf{Proof:} We follow the arguments from theorem~2 in~\cite{bmrv}
(preserving the notation). 
Consider any randomized scheme with two-parts memory. 
Denote by $C$ the cached memory (of size $\poly(\log m)$) and
by $B$ the main part of the memory of size $s$ 
(the scheme answers queries  with one bit probe to $C$). 
Our aim is to prove a lower bound for  $s$.

The proof is based on the bound for the size of cover-free families of sets
proven by Dyachkov an Rykov \cite{dr}.
Let us remind that a family of sets $F$ is called called $r$-cover-free
if $f_0\not\subseteq f_1\cup\ldots\cup f_r$ for all distinct
$f_0,\ldots,f_r\in F$.  

First we take a large enough $\frac{1}{\varepsilon}$-cover free family $F$ 
of sets of size $n$ from the universe $\{1,\ldots,m\}$. The construction from
\cite[theorem~3.1]{erdos} guarantees that there exists a family $F$ such that
 $$
 |F|\ge \frac{{m\choose {\varepsilon n} }}{{{ n\choose{\varepsilon n} } }^2}
 =2^{\varepsilon n\log\frac{m}{n^2} + O(\log m)}.
 $$
By assumption, each  set  $f\in F$ can  be represented in our scheme by 
some pair $(B,C)$ (the main storage and the cached memory).
Notice that 
$$|F|=2^{\varepsilon n\log\frac{m}{n^2} + O(\log m)} \gg 2^{|B|}=2^{\poly(\log m)}.$$ 
Hence, the exists
some value $C$  and some $F'\subset F$
of size
 $$
 |F'|\ge |F|/ 2^{\poly(\log m)}
 =2^{\om(\varepsilon n\log m)}.
 $$
such that all sets from the family  $F'$ share in our scheme
the same value  $C$ of the cached memory. Further,
we repeat word for word the proof of theorem~2 from~\cite{bmrv}
with substitute $F'$ instead of $F$.

\section{Conclusion.}

In this paper we constructed an effective probabilistic bit-probe scheme with one-sided error.
The used space is close to the trivial information-theoretic
 lower bound $\om(n\log m)$. 
The scheme answers queries  ``$x\in A$?'' with a small
one-sided  error and requires only poly-logarithmic (in the size of the universe)
cached memory and  \emph{one bit} (\emph{two bits} in the version with effective encoding) 
from the main part of the  memory.
 
For reasonable values of parameters (for $n\gg \poly(\log m)$) the size of 
our scheme $O(n\log^2 m)$ with a cache of size $\poly(\log m)$ is 
below the lower bound $\om(n^2\log m)$ proven in~\cite{bmrv}
for one-probe schemes with one-sided errors without cached data dependent on $A$.
The gap between our upper bounds and the trivial lower bound is a $(\log m)$-factor.

The following  questions remain open: 
How to construct a bit-probe memory scheme with one-sided
error and effective encoding and decoding that requires to read only \emph{one} bit from the
main part of memory to answer queries? What is the minimal size of the cached memory
required for  a bit-probe scheme with one-sided error, with
space of size $O(n\log m)$?
 
\bigskip

\noindent
The author thanks Daniil~Musatov for useful discussions, and anonymous referees 
of CSR2011 for deep and very helpful comments.

\end{document}